\documentclass[twocolumn,twoside,slac_two]{revtex4}
\usepackage{graphicx}
\usepackage{fancyhdr}
\usepackage{longtable}
\usepackage{amsmath}
\begin{document}

\title{Measurements of the CKM Angle ${\boldsymbol \alpha}$ at BaBar}

\author{S. Stracka on behalf of the BaBar Collaboration}
\affiliation{Universit\`a degli Studi di Milano and INFN, Sezione di Milano - I-20133 Milano, Italy}

\begin{abstract}

We present improved measurements of the branching fractions and $CP$-asymmetries
in the $B^0\to\pi^+\pi^-$, $B^0\to\pi^0\pi^0$, and $B^+\to\rho^+\rho^0$ decays, 
which impact the determination of $\alpha$.
We find 
\begin{eqnarray*}
S_{\pi\pi}^{+-} &=& -0.68 \pm 0.10 \pm 0.03 \\
C_{\pi\pi}^{+-} &=& -0.25 \pm 0.08 \pm 0.02 \\
C_{\pi\pi}^{00} &=& -0.43 \pm 0.26 \pm 0.05 \\
{\cal B}(B^0\to \pi^0\pi^0) &=& (1.83 \pm 0.21 \pm 0.13)\times 10^{-6}
\end{eqnarray*}
for $B\to\pi\pi$ decays, and 
\begin{eqnarray*}
{\cal B}(B^+\to \rho^+\rho^0) &=& (23.7 \pm 1.4 \pm 1.4)\times 10^{-6} \\
f_L(\rho^+\rho^0) &=& 0.950\pm 0.015 \pm 0.006\\
\alpha_{\rho\rho} &=& (92.4^{+6.0}_{-6.5})^{\circ} 
\end{eqnarray*}
for $B\to \rho \rho$ decays.

The combined branching fractions of 
$B\to K_1(1270)\pi$ and $B\to K_1(1400)\pi$ decays 
are measured for the first time and allow a novel determination 
of $\alpha$ in the $B^0\to a_1(1260)^{\pm}\pi^{\mp}$ decay channel.
We obtain
\begin{eqnarray*}
{\cal B}(B^0\to K_1(1270)^+\pi^- + K_1(1400)^+\pi^-) &=& (3.1^{+0.8}_{-0.7})\times 10^{-5} \\
{\cal B}(B^+\to K_1(1270)^0\pi^+ + K_1(1400)^0\pi^+) &=& (2.8^{+2.9}_{-1.7})\times 10^{-5} \\
\alpha_{a_1\pi} &=&(79\pm 7\pm 11)^{\circ} .
\end{eqnarray*}

These measurements are performed using the final dataset collected by the BaBar 
detector at the PEP-II B-factory.

\end{abstract}

\maketitle

\thispagestyle{fancy}

\section{Introduction}

The primary goal of the experiments based 
at the $B$ factories is to test the Cabibbo-Kobayashi-Maskawa 
(CKM) picture of $CP$ violation in the standard model of electroweak 
interactions~\cite{CKM}. 
This can be achieved by measuring the angles and sides of the 
Unitarity Triangle in a redundant way.

An effective value $\alpha_{\rm eff}$ for the CKM phase 
$\alpha\equiv \arg(-V^{}_{td}V^*_{tb}/V^{}_{ud}V^*_{ub})$ 
can be extracted from the time-dependent analysis of $B$ meson decays 
dominated by tree-level $b\to u\bar{u}d$ amplitudes, such as 
$B^0 \to \pi^+\pi^-$, $B^0 \to \rho^+\rho^-$, $B^0 \to \rho^{\pm}\pi^{\mp}$, 
and $B^0 \to a_1(1260)^{\pm}\pi^{\mp}$. 
The current average values of $\alpha$, 
$\alpha=(92 \pm 7)^{\circ}$~\cite{UTFIT} and 
$\alpha=(89^{+4.4}_{-4.2})^{\circ}$~\cite{CKMFITTER}, obtained with different
statistical techniques, 
are based solely 
on the analysis of $B\to\pi\pi$, $B\to\rho\rho$, and $B\to \rho\pi$ 
decays.

The measurement of the angle $\alpha$ 
has witnessed significant progress over the past year.
The following sections are organized as follows: 
a brief introduction on the experimental technique is given in 
Sec.~\ref{sec:exptech}; the summer 2008 update of  
the measurement of the time-dependent $CP$-violating asymmetries in 
$B^0\to \pi^+\pi^-$ decays and of the branching fractions (BFs) of 
$B^0\to\pi^0\pi^0$ decays is reported in Sec.~\ref{sec:btohh};
Sec.~\ref{sec:btorhorho} describes the 2009 update of the 
BF measurement of $B^+\to\rho^+\rho^0$ decays, 
and its impact on the precision of the determination of $\alpha$;
in Sec.~\ref{sec:btoa1pi}, we introduce the first measurement 
of $B\to K_1(1270)\pi$ and $B\to K_1(1400)\pi$ decays and a new 
determination of $\alpha$ in $B^0\to a_1(1260)^{\pm}\pi^{\mp}$ decays.

\section{Experimental techniques}
\label{sec:exptech}

The interference between the direct tree decay (which carries 
the weak phase $\gamma$) and decay after $B^0\bar{B}^0$ mixing
(which carries a weak phase $2\beta$) results in a time-dependent 
decay-rate asymmetry that is sensitive to the angle 
$2\beta + 2\gamma = 2\pi - 2\alpha$.

At the asymmetric-energy $e^+e^-$ B-factory PEP-II, running at 
a center of momentum (CM) energy of $10.58\,{\rm GeV}$, 
a $B\bar{B}$ pair is coherently produced in the decay 
of a $\Upsilon(4S)$ resonance.
The resulting $B\bar{B}$ system has a boost $\beta\gamma\approx 0.56$  
with respect to the laboratory frame.
By means of this experimental device it is possible to measure the 
decay vertex displacement $\Delta z$ of the two $B$ mesons in the event, 
and hence their proper-time difference 
$\Delta t_{meas}\approx \frac{\Delta z}{\beta\gamma c}$. 

One of the $B$ mesons ($B_{rec}$) 
is fully reconstructed according to the final state of interest.
In order to study the time-dependence of the decay rates, 
it is necessary to measure the proper-time difference $\Delta t$ 
between the two $B$ mesons in the event and to identify the 
flavor of the other $B$-meson ($B_{tag}$).
The flavor and the decay vertex
position of $B_{tag}$ are therefore identified from its decay products.

The decay-rate distribution for $B^0$ ($\bar{B}^0$)
decays to a $CP$-eigenstate, such as $\pi^+\pi^-$, is given by:
\begin{eqnarray}
\frac{d N}{d \Delta t} &=& \frac{e^{-|\Delta t|/\tau}}{4\tau}
\left\{ 1 - q_{tag}\left[ C \cos(\Delta m_d \Delta t) 
\phantom{\frac{e^{1}}{4}} \right. \right. \label{eq:rateasym} \\
 & & \phantom{\frac{e^{-|\Delta t|/\tau}}{4\tau}
\left\{ 1 - q_{tag}\left[\right.\right.} - S \sin(\Delta m_d \Delta t)\left]\left\},\phantom{\frac{e^{1}}{4}}\right.\right. \nonumber
\end{eqnarray}
where $\tau = (1.536 \pm 0.014)\,\rm{ps}$~\cite{PDG} is the mean $B$ lifetime, 
$\Delta m_d = (0.502 \pm 0.007)\,\rm{ps}^{-1}$ is the 
$B^0-\bar{B}^0$ mixing frequency~\cite{PDG},
and $q_{tag}=+1$ ($-1$) if the $B_{tag}$ decays as a $B^0$ (${\bar B}^0$).
The parameters $S$ and $C$ describe mixing-induced and direct 
$CP$-violation, respectively, and are defined as:
\begin{eqnarray*}
S=\frac{2{\rm Im}\lambda}{1+|\lambda|^2},  & &
C=\frac{1-|\lambda|^2}{1+|\lambda|^2},  
\end{eqnarray*}
with $\lambda = \frac{q}{p}\frac{\bar A}{A}$, where $q/p$ is 
related to the $B^0-\bar{B}^0$ mixing, and $A$ ($\bar A$) is 
the amplitude of the decay of a $B^0$ ($\overline{B}^0$)
to the final state under study.
If only the tree amplitude contributes 
to the decay, $S=\sin (2\alpha)$ and $C=0$. 
However,  $b\to u\bar{u}d$ transitions receive sizeable contributions 
from penguin (loop) amplitudes, which carry different strong and weak phases.
This contribution can result in non-zero direct $CP$-violation $C\neq 0$ and 
modifies $S$ into 
\begin{equation}
\label{salphaeff}
S=\sin(2\alpha_{\rm eff}) \sqrt{1-C^2}.
\end{equation}
The angle $\alpha_{\rm eff}$ coincides with $\alpha$ in the limit of 
vanishing penguin contributions.
In order to constrain $\Delta\alpha\equiv \alpha - \alpha_{\rm eff}$, 
techniques based on the SU(2) isospin 
symmetry (for decays to a $CP$-eigenstate, such as 
$B^0\to \pi^+\pi^-, \rho^+\rho^-$) or the SU(3) approximate 
flavor symmetry (for decays to a non $CP$-eigenstate, such as 
$B^0\to \rho^{\pm}\pi^{\mp}, a_1(1260)^{\pm}\pi^{\mp}$) have been devised, and 
are discussed in the remaining of this paper.

A neural network based tagging algorithm~\cite{TAGGING} 
is used to determine whether the $B_{tag}$ is a $B^0$ or a $\bar{B}^0$. 
Events are separated according to the particle content 
of the $B_{tag}$ final state into events where 
there are leptons, kaons and pions, for a total of seven mutually exclusive 
categories. 
The performance of the tagging algorithm is characterized by 
the efficiency $\epsilon_{\rm tag}$ in the determination of the flavor 
of $B_{tag}$ and by the mistag probability $\omega$, and depends on the 
tagging category. 
The $\Delta t$ distribution of Eq.~\ref{eq:rateasym} is convolved with 
a detector resolution function, which differs for signal and background, and 
is parameterized as a triple Gaussian. Dilution from incorrect assignment 
of the flavor of $B_{tag}$ is also taken into account:
\begin{eqnarray}
\frac{d N}{d \Delta t_{meas}} & = & \frac{e^{-|\Delta t|/\tau}}{4\tau} \times 
\left\{ 1 - q_{tag}\Delta \omega - \phantom{\frac{e^{1}}{4}} \right. \\ \nonumber
& & q_{tag}(1-2\omega)\left[ C \cos(\Delta m_d \Delta t) - \phantom{\frac{e^{1}}{4}}  \right.\\ \nonumber
& & \phantom{q_{tag}(1-2\omega)\left[\right.} S \sin(\Delta m_d \Delta t)\left]\left\} \phantom{\frac{e^{1}}{4}}\right.\right. \\ \nonumber
& & \otimes  R(\Delta t_{meas}-\Delta t), \nonumber
\label{eq:rateexp}
\end{eqnarray}
where $(1-2\omega)$ is the dilution factor, $\Delta \omega$ is the 
difference in mistag probabilities 
$\Delta\omega \equiv \omega_{B^0}- \omega_{\bar B^0}$
and $R$ is the resolution function. 
The parameters of the resolution function are obtained from a fit to a 
large sample of fully reconstructed $B$ decays, as in \cite{TAGGING}, 
and are free to differ between tagging categories.

The analyses of the two-body and quasi-two-body decays described in the 
remaining of this paper rely 
on a common strategy for the suppression of the continuum $e^+e^-\to q\bar{q}$ 
background ($q=u,d,s,c$), which represents the most abundant 
source of background.
Two kinematic variables, the energy substituted mass $m_{ES}=\sqrt{s/4-p^2_B}$ 
and the energy difference $\Delta E=E_B-\sqrt{s}/2$, where $\sqrt{s}$ is 
the $e^+e^-$ CM energy and the 
four-momentum $(E_B,p_B)$ of the $B$ meson is defined in the CM frame, 
allow to discriminate correctly reconstructed $B$ candidates 
(for which the distribution of $m_{ES}$ peaks at the $B$-meson mass and that 
of $\Delta E$ peaks at zero) and fake candidates resulting from random 
combination of particles (for which $m_{ES}$ follows a phase-space 
distribution and $\Delta E$ is approximately flat).
Topological variables provide further distinction between the jet-like shape of 
continuum events and the more isotropic $B$ decays,
and can be  combined into multivariate classifiers, such as neural network
and Fisher discriminant, to enhance the discriminating power.
The signal and background yields and $CP$ asymmetries 
are extracted via an extended unbinned maximum-likelihood (ML) fit to
the data.

\section{Isospin analysis of $\boldsymbol{B\to \pi\pi}$ decays}
\label{sec:btohh}

\subsection{$\boldsymbol{B^0\to \pi^+\pi^-}$}

In the $\pi\pi$ system the penguin pollution is greatest. 
The tree ($T$) and penguin ($P$) amplitudes each contribute, with different 
weak ($\phi$) and strong ($\delta$) phases, with comparable magnitude.
Direct $CP$ violation, which is given by 
$A_{CP}=2\sin\phi\sin\delta/(|T/P|+|P/T|+2\cos\phi\cos\delta)$, 
can, therefore, be within observational reach.

\begin{figure}[t]
\vskip 3mm
\includegraphics[width=80mm]{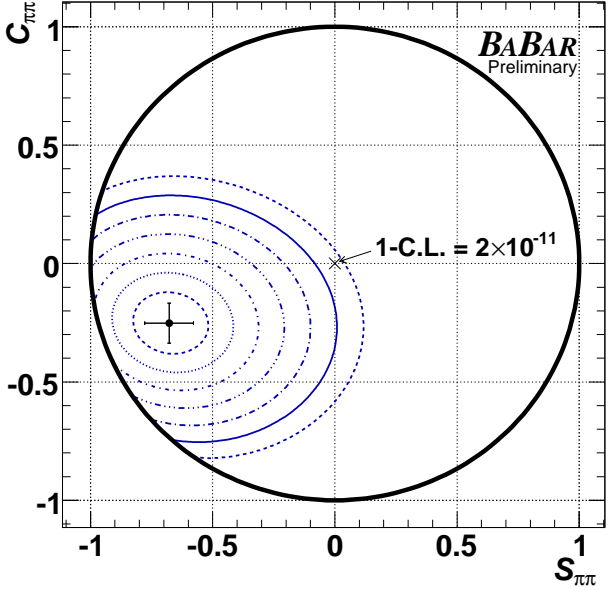}%
\caption{\label{btopipiCS} $S_{\pi\pi}^{+-}$ and $C_{\pi\pi}^{+-}$ in $B^0\to \pi^+\pi^-$: 
the central values, errors, and confidence-level (CL)
contours, calculated from the square root of the change in the value 
of $-2 \ln {\cal L}$ compared with its value at the minimum~\cite{BTOPIPI}. 
The systematic errors are included. The
measured value is $6.7\sigma$ from the point of no CP violation ($S_{\pi\pi}^{+-}=0$ and $C_{\pi\pi}^{+-}=0$).
}
\end{figure}

$B^0\to\pi^+\pi^-$ decays are analyzed with the full BaBar 
dataset of $467 \pm 5$ million $B\overline{B}$ pairs~\cite{BTOPIPI}. 
A simultaneous ML fit to the $\pi^+\pi^-$, $\pi^+K^-$, $K^+\pi^-$, 
and $K^+K^-$ final states is performed. 
$K-\pi$ separation is obtained by 
particle-identification (PID) observables (the Cherenkov angle $\Theta_C$ in
the DIRC~\cite{DIRC} 
and ionization-energy loss $dE/dx$ in the tracking devices~\cite{NIM}).
Additional separation between the final states under study is achieved 
from $\Delta E$: since the $B$ meson is reconstructed from two 
oppositely charged tracks that are both given the pion mass hypothesis, 
each charged $K$ in the final state results in a $\Delta E$ displacement 
of about $-45~{\rm MeV}$.
We extract $1394\pm 54$ signal events.
From the time distribution of $B^0\to \pi^+\pi^-$ decays  
a non-zero mixing-induced $CP$ violation asymmetry 
$S_{\pi\pi}^{+-}=-0.68\pm 0.10\pm 0.03$ is observed with significance 
$6.3\sigma$~\cite{BTOPIPI}, as shown in Fig.~\ref{btopipiCS}. 
A non-zero direct $CP$ violation asymmetry 
$C_{\pi\pi}^{+-}=-0.25\pm 0.08\pm 0.02$ is
also extracted with significance $3.0\sigma$~\cite{BTOPIPI}.  

\subsection{$\boldsymbol{B^0\to \pi^0\pi^0}$}
The $B^0\to \pi^0\pi^0$ decay is formed from pairs of 
$\pi^0\to \gamma\gamma$ candidates, where one of the photons can
eventually be reconstructed from two tracks coming from a photon conversion 
$\gamma \to  e^+e^-$ inside the electromagnetic calorimeter.

The yield and the flavor tag- and time-integrated $CP$ asymmetry 
$A_{CP}^{00}=-C_{\pi\pi}^{00}$ are 
obtained from a ML fit to the kinematic variables $\Delta E$ and $m_{ES}$ and 
the output of a neural network $NN$ computed from event-shape variables, as 
well as the output of the $B$-flavor tagging algorithm.
The background model accounts for correlations between $NN$ and $m_{ES}$.
We observe $247\pm 29$ signal events 
(corresponding to 
${\cal B}(B^0\to \pi^0\pi^0)=(1.83 \pm 0.21 \pm 0.13)\times 10^{-6}$) and 
extracts $C_{\pi\pi}^{00}=-0.43\pm 0.26 \pm 0.05$~\cite{BTOPIPI}.
Since no reliable vertex information is extracted, $S_{\pi\pi}^{00}$ can 
not be determined.

\subsection{Isospin analysis of $\boldsymbol{B\to \pi\pi}$ decays}
\begin{figure}[h]
\includegraphics[width=80mm]{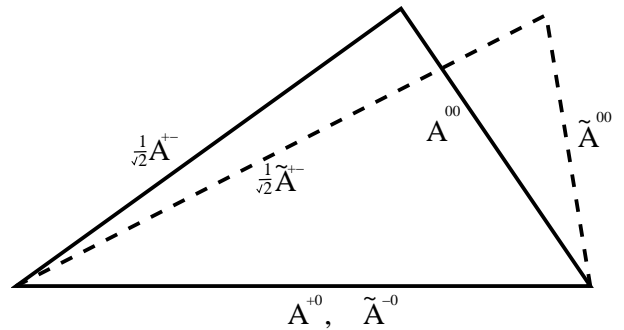}%
\caption{\label{isospin} 
Triangles in the complex plane describing the isospin relations Eq.\,(\ref{su2a}) and Eq.\,(\ref{su2b}).}
\end{figure}

The rates and $CP$ asymmetries of $B^0\to \pi^+\pi^-$ and 
$B^0\to \pi^0\pi^0$ decays are combined with the results for the 
$B^+\to \pi^+\pi^0$ mode in a model-independent isospin analysis~\cite{ISOSPINPIPI}. 
Under the isospin symmetry, $B\to \pi\pi$ amplitudes can be decomposed 
in isospin $I=0$ ($A_0$) and $I=2$ ($A_2$) amplitudes. By virtue of 
Bose statistics, $I=1$ contributions are forbidden.
The following relations hold~\cite{ISOSPINPIPI}:
\begin{equation}
(1/\sqrt{2})A^{+-} = A_2-A_0,
\end{equation}
\begin{equation}
A^{00}=2A_2+A_0, ~ A^{+0}=3A_2, 
\end{equation}
where $A^{ij}$ ($\overline{A}^{ij}$) are the amplitudes of $B$ ($\overline{B}$)
decays to the $\pi^i\pi^j$ final state. 
This yields the complex triangle relations:
\begin{equation}
\label{su2a}
\frac{1}{\sqrt{2}}A^{+-} = A^{+0}-A^{00},
\end{equation}
\begin{equation}
\label{su2b}
\frac{1}{\sqrt{2}}\overline{A}^{+-} = \overline{A}^{-0}-\overline{A}^{00}.
\end{equation}
Tree amplitudes receive contributions from both $A_0$ and $A_2$, 
while gluonic penguin diagrams are pure $I=0$ amplitudes 
and do not contribute to $B^+\to\pi^+\pi^0$ amplitudes.
Possible contributions from electroweak penguins (EWP), 
which do not obey SU(2) 
isospin symmetry, are assumed to be negligible and are therefore ignored. 
Under this assumption, $|A^{+0}|=|\overline{A}^{-0}|$ (a sizeable 
contribution from EWPs would result in $|A^{+0}|\neq|\overline{A}^{-0}|$
and would be signalled by an evidence of
direct $CP$ violation in $B^+\to \pi^+\pi^0$ decays). 
If $A^{+0}$ and $\overline{A}^{-0}$ are aligned with a suitable choice of 
phases, the relations (\ref{su2a}) and (\ref{su2b}) 
can be represented in the complex plane by two triangles (Fig.~\ref{isospin}),
and the phase difference between $A^{+-}$ and $\overline{A}^{+-}$ 
is $2\Delta\alpha$.

Constraints on the CKM angle $\alpha$ and on the penguin contribution 
$\Delta \alpha$ 
are obtained from a confidence level (CL) scan over the parameters of interest, 
$\alpha$ and $|\Delta \alpha|$.
Assuming the isospin-triangle relations (\ref{su2a}) and (\ref{su2b})
and the expression (\ref{salphaeff}), a 
$\chi^2$ for the five amplitudes $(A^{+0}, A^{+-}, A^{00}, 
\overline{A}^{+-}, \overline{A}^{00})$ is calculated from the 
measurements summarized in Table~\ref{inputpisu2}, and minimized 
with respect to the parameters that don't enter the scan.
The $1-CL$ values are then calculated from the probability of 
the minimized $\chi^2$.

\begin{table}[t]
\begin{center}
\caption{Summary of the input to the isospin analysis of the $\pi\pi$ system~\cite{BTOPIPI,BPTOPIPI}.}
\begin{tabular}{|l|c|c|}
\hline 
\bf{Mode} & $\boldsymbol{{\cal B}(\times 10^{-6})}$ & $\boldsymbol{C}$ \\
\hline
$\pi^+\pi^-$ & $5.5\phantom{0}  \pm 0.4\phantom{0}  \pm 0.3\phantom{0} $ & $-0.25 \pm 0.08 \pm 0.02$   \\
$\pi^+\pi^0$ & $5.02 \pm 0.46 \pm 0.29$ & ($-0.03 \pm 0.08 \pm 0.01$) \\
$\pi^0\pi^0$ & $1.83 \pm 0.21 \pm 0.13$ & $-0.43\pm 0.26 \pm 0.05$    \\
\hline
\end{tabular}
\label{inputpisu2}
\end{center}
\end{table}

\begin{figure}[b]
\includegraphics[width=80mm]{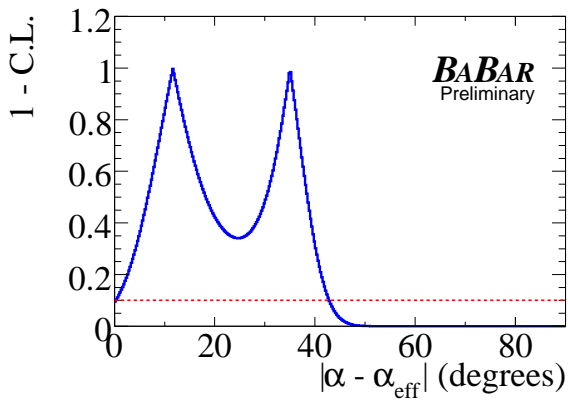}%
\caption{\label{deltaalphapipi} 
Projection of the $1-CL$ scan on $\Delta \alpha$ for the $\pi\pi$ 
system~\cite{BTOPIPI}.}
\end{figure}

\begin{figure}[b]
\includegraphics[width=80mm]{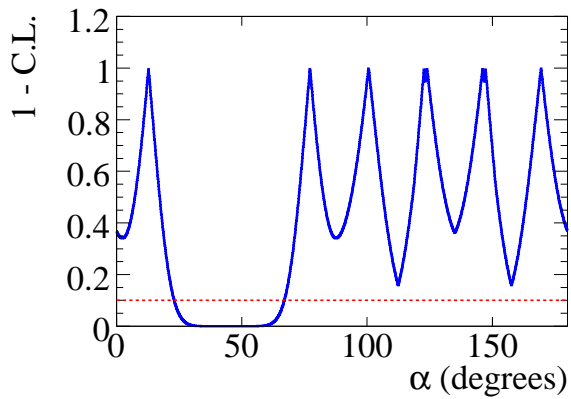}%
\caption{\label{alphapipi} 
Projection of the $1-CL$ scan on $\alpha$ for the $\pi\pi$ 
system~\cite{BTOPIPI}.}
\end{figure}

The results of the isospin analysis 
are shown in Fig.~\ref{deltaalphapipi} and Fig.~\ref{alphapipi}. 
$\Delta\alpha$ is extracted with a four-fold ambiguity, 
which can be graphically represented as a flip of either triangle 
around $A^{+0}$.
An additional two-fold ambiguity arises from the trigonometric relation 
$S_{\pi\pi}^{+-}=\sin(2\alpha_{\rm eff}) \sqrt{1-C_{\pi\pi}^{+-2}}$.
This results in a global eight-fold ambiguity in the 
range $[0,180]^{\circ}$ on the extraction of $\alpha$.
A value $\Delta\alpha < 43^{\circ}$ at 90\% CL is
obtained, which dominates the uncertainty on $\alpha$~\cite{BTOPIPI}.
Considering only the solution consistent with the results of global CKM
fits, $\alpha$ is in the range $[71, 109]^{\circ}$ at the 
68\% CL~\cite{BTOPIPI}.


The limiting factor in the extraction of $\Delta \alpha$ is the knowledge 
of $|A^{00}|$ and $|\overline{A}^{00}|$, which is severely limited by the 
available statistics.
A significant increase in statistics is therefore 
required in order to perform a precision measurement of $\alpha$ in this 
channel. 
A measurement of $S_{\pi\pi}^{00}$, which  
would aid resolving some ambiguities on $\alpha$, can only be addressed 
with Super B factory luminosities~\cite{SUPERB}.

\section{Isospin analysis of $\boldsymbol{B\to \rho\rho}$ decays}
\label{sec:btorhorho}
With respect to $B\to \pi\pi$ decays, $B\to \rho\rho$ decays 
have a more favourable penguin to tree amplitude ratio.
Moreover, the BF for $B^0\to \rho^+\rho^-$ decays  
is greater than that for $B^0\to \pi^+\pi^-$ decays by a factor of 
$\approx 5$~\cite{BTORHOPRHOM}.
Finally, the $B^0\to \rho^0\rho^0$ decay can be reconstructed from a final 
state 
consisting of all charged tracks, with enough efficiency to 
allow for a measurement of $S_{\rho\rho}^{00}$ with the present 
statistics~\cite{BTORHO0RHO0}.
Despite these many advantages with respect to the isospin analysis of  
$\pi\pi$ decays, the $\rho\rho$ system exhibits some potential 
complications.

\begin{figure}[h]
\includegraphics[width=80mm]{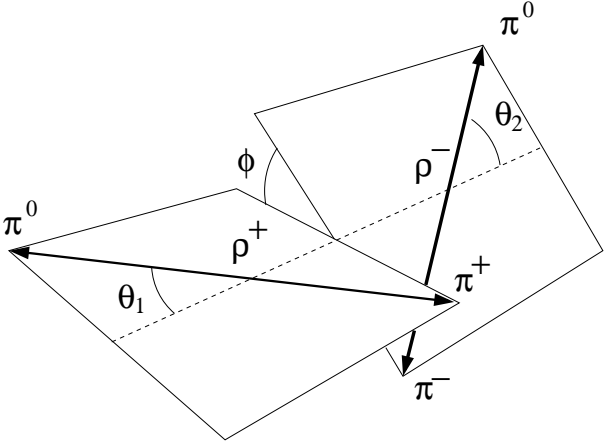}
\caption{\label{angular} 
Definition of the $\theta_1$ and $\theta_2$ angles in $B\to \rho\rho$ 
decays~\cite{BTORHOPRHOM}.}\end{figure}

In $B^0\to \rho^+\rho^-$ transitions, a pseudo-scalar particle 
decays into two vector mesons. Three helicity states ($H=0,\pm 1$),
with different $CP$ transformation properties, can therefore contribute 
to the decay~\cite{KAGAN}. The $H=0$ state corresponds to longitudinal 
polarization 
and is $CP$-even, while the transverse polarization states $H=+1$ and $H=-1$ 
(which are superpositions of S-, P-, and D-wave amplitudes) 
have not a definite $CP$-eigenvalue.
Isospin relations similar to Eq. (\ref{su2a}) and (\ref{su2b})
hold separately for each polarization state.
The analysis of the angular distribution of $B^0\to \rho^+\rho^-$ decays 
allows to determine the longitudinal polarization fraction $f_L$:
\begin{eqnarray}
\frac{1}{\Gamma}\frac{d^2\Gamma}{d\cos\theta_1 d\cos\theta_2}&\propto &
4 f_L \cos^2\theta_1\cos^2\theta_2 \nonumber \\
& & + (1-f_L)\sin^2\theta_1\sin^2\theta_2,
\end{eqnarray}
where $\theta_1$ ($\theta_2$) is the angle between the daughter $\pi^0$
and the direction opposite to the $B$
direction in the $\rho^+$ ($\rho^-$) rest frame, as shown in Fig.~\ref{angular}.
Since experimental measurements have shown the decay to be dominated 
by the longitudinal, $CP$-even polarization, it is not necessary to separate
the definite-$CP$ contributions of the transverse polarization by means of 
a full angular analysis.

A second complication arises because the $\rho$ mesons have finite width, 
thus allowing for the two $\rho$ mesons in the decay to have different masses.
Since the Bose-Einstein symmetry does not hold, 
the wave function of the $\rho\rho$ system can be 
anti-symmetric, and isospin $I=1$ amplitudes are allowed, breaking 
the isospin relations Eq.\,(\ref{su2a}) and (\ref{su2b})~\cite{FALK}.
The stability of the fitted $CP$-violation parameters against 
the restriction of the $\pi\pi$ invariant mass window used to select the
$\rho$ candidates shows however that possible isospin violation effects are 
below the current sensitivity. 

\subsection{$\boldsymbol{B^+\to \rho^+\rho^0}$}

\begin{table*}[t]
\begin{center}
\caption{Summary of the input to the isospin analysis of the $\rho\rho$ 
system~\cite{BTORHOPRHOM,BTORHO0RHO0,BTORHORHO}.}
\begin{tabular}{|l|c|c|c|c|}
\hline 
\bf{Mode} & $\boldsymbol{{\cal B}(\times 10^{-6})}$ & $f_L$ & $\boldsymbol{C}$& $\boldsymbol{S}$ \\
\hline
$\rho^+\rho^-$ & $25.5  \pm 2.1  ^{+3.6}_{-3.9} $ & $0.992 \pm 0.024 ^{+0.026}_{-0.013}$ & $0.01\pm 0.15 \pm 0.06$ & $-0.17\pm 0.20^{+0.05}_{-0.06}$   \\
$\rho^+\rho^0$ & $23.7 \pm 1.4 \pm 1.4$ & $0.950 \pm 0.042 \pm 0.006$ & ($0.054 \pm 0.055 \pm 0.010$) & - \\
$\rho^0\rho^0$ & $0.92 \pm 0.32 \pm 0.14$ & $0.75^{+0.11}_{-0.14}\pm 0.04$ & 
$0.2\pm 0.8 \pm 0.3$    &$0.3\pm 0.7 \pm 0.2$    \\
\hline
\end{tabular}
\label{inputrhosu2}
\end{center}
\end{table*}

The $B^+\to \rho^+\rho^0$ decay analysis has been updated using the 
final BaBar dataset of $424\,\rm{fb}^{-1}$~\cite{BTORHORHO}, superseding the previous 
analysis based on $211\,\rm{fb}^{-1}$~\cite{BTORHORHOOLD}. 
An analysis of the angular distributions of $B^+\to \rho^+\rho^0$ decays
is performed. The signal yield and longitudinal polarization fraction is 
extracted via a ML fit to the kinematic quantities $m_{ES}$, $\Delta E$, 
the output of a neural network $NN$ based on event-shape variables, 
the mass of the $\rho^+$ and $\rho^0$ candidates, and the cosines of 
the helicity angles $\theta_{\rho^+}$ and $\theta_{\rho^0}$, where 
$\theta_{\rho^+}$ ($\theta_{\rho^0}$) is the angle between the daughter $\pi^0$
($\pi^-$) and the direction opposite to the $B$
direction in the $\rho^+$ ($\rho^0$) rest frame.
\begin{figure}[h]
\includegraphics[width=80mm]{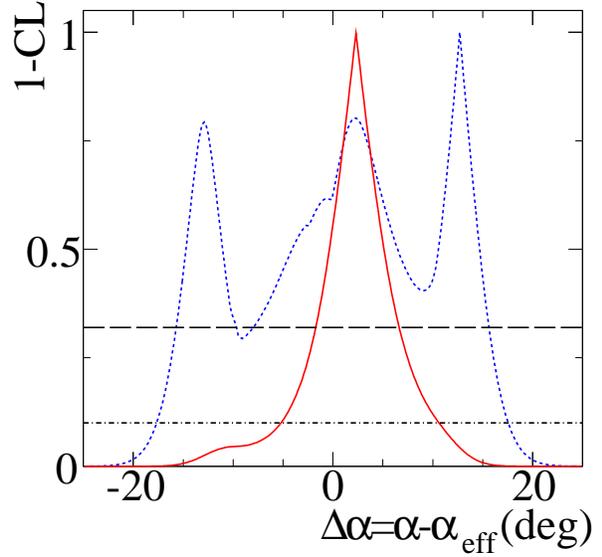}
\caption{\label{deltaalpharhorho} 
Projection of the $1-CL$ scan on $\Delta \alpha$ for the $\rho\rho$ 
system~\cite{BTORHORHO}.}
\end{figure}
\begin{figure}[h]
\includegraphics[width=80mm]{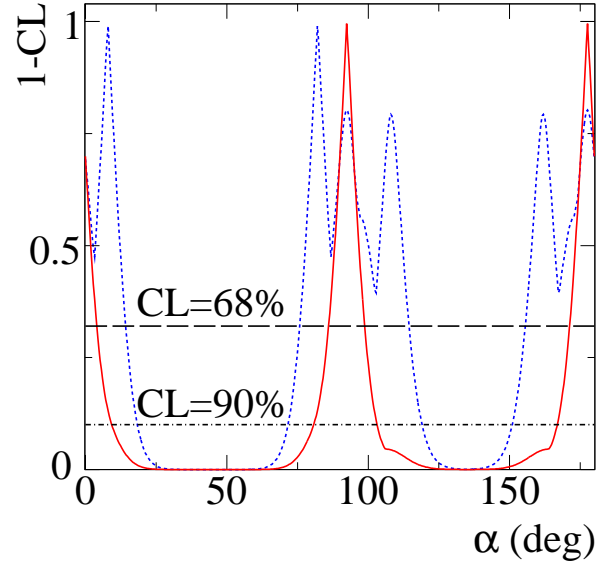}
\caption{\label{alpharhorho} 
Projection of the $1-CL$ scan on $\alpha$ for the $\rho\rho$ 
system~\cite{BTORHORHO}.}
\end{figure}
Improvements have been introduced on the charged particle reconstruction 
and on the background model, which takes into account correlations 
between $NN$, the cosine of the helicity angle, 
and the $\pi\pi$ invariant mass for each $\rho$ meson in the final state.
The measured BF increases from 
$(18.2 \pm 3.0) \times 10^{-6}$~\cite{BTORHORHOOLD} up to 
$(23.7 \pm 1.4 \pm 1.4) \times 10^{-6}$~\cite{BTORHORHO}. 
The longitudinal polarization fraction
is $f_L=0.950\pm 0.015 \pm 0.006$~\cite{BTORHORHO}.  
The measured direct $CP$-violation asymmetry 
$A_{CP}\equiv \frac{\Gamma(B^-\to\rho^-\rho^0) - \Gamma(B^+\to\rho^+\rho^0)}
{\Gamma(B^-\to\rho^-\rho^0) + \Gamma(B^+\to\rho^+\rho^0)}$ is 
$A_{CP}=-0.054\pm 0.055\pm 0.010$, which is consistent with 0. 
This result indicates that the contribution from EWPs is negligible, and 
the isospin analysis holds within an uncertainty of 
$1-2^{\circ}$~\cite{ISOBREAKING}.

The BFs, longitudinal polarization fractions, direct and mixing-induced 
$CP$ violation asymmetries for $B\rightarrow \rho\rho$ decays 
are used as input to the isospin analysis, and are summarized in 
Table~\ref{inputrhosu2}.
The BF's of $B^+\rightarrow \rho^+\rho^0$ and 
$B^0\rightarrow \rho^+\rho^-$ are now very similar and much higher 
than that for the $B^0\rightarrow \rho^0\rho^0$ penguin transition.
As a consequence, the isospin triangles do not close, {\em i.e.} 
$|A^{+-}|/\sqrt{2}+|A^{00}|<|A^{+0}|$. This results in a degeneracy of the
eight-fold ambiguity on $\alpha$ into a four-fold ambiguity, corresponding
to peaks in the vicinity of $0^{\circ}$, $90^{\circ}$ (two degenerate peaks), 
$180^{\circ}$, as shown in Fig.~\ref{deltaalpharhorho} and 
Fig.~\ref{alpharhorho}. 
A value $-1.8^{\circ}<\Delta\alpha < 6.7^{\circ}$ at 68\% CL is
obtained.
Considering only the solution consistent with the results of global CKM
fits, $\alpha=92.4^{+6.0}_{-6.5}$. The precision on 
$\alpha$ is now at the level of $5\%$.

\section{$\boldsymbol{B^0\to a_1(1260)^{\pm}\pi^{\mp}}$}
\label{sec:btoa1pi}
It is possible to extract $\alpha$ from $B$ decays to 
final states that are not $CP$-eigenstates~\cite{ALEKSAN}, 
such as $B^0\to a_1(1260)^{\pm}\pi^{\mp}$ decays.
The relevant amplitudes are:
\begin{eqnarray}
A_+ \equiv A(B^0\to a_1^+\pi^-), & \overline{A}_+ \equiv A(\bar{B}^0\to a_1^-\pi^+), \\ 
A_- \equiv A(B^0\to a_1^-\pi^+), & \overline{A}_- \equiv A(\bar{B}^0\to a_1^+\pi^-).  
\end{eqnarray}
The time distribution for this decay mode is given by:
\begin{eqnarray}
\nonumber
\frac{d N^{a_1^{\pm}\pi^{\mp}}}{d \Delta t_{meas}} & = & 
(1\pm A_{CP})\frac{e^{-|\Delta t|/\tau}}{4\tau} \left\{ 1 - q_{tag}\Delta \omega + \phantom{\frac{e^{1}}{4}} \right. \\ \nonumber
& &  q_{tag}(1-2\omega)\left[ (S \pm \Delta S)\sin(\Delta m_d \Delta t)- 
\phantom{\frac{e^{1}}{4}}\right. \\ \nonumber
& & \phantom{q_{tag}(1-2\omega)\left[\right.} (C \pm \Delta C)\cos(\Delta m_d \Delta t) \left]\left\}
\phantom{\frac{e^{1}}{4}}\right.\right.\\ \nonumber
& &  \otimes  R(\Delta t_{meas}-\Delta t),
\label{eq:rateexpa1}
\end{eqnarray}
where $A_{CP}$ is the time- and flavor-integrated charge asymmetry, and 
\begin{eqnarray}
S\pm \Delta S & \equiv & \frac{2{\rm Im} \left(e^{-2i\beta}\overline{A}_{\mp}A^{*}_{\pm}\right)}{|A_{\pm}|^2+|\overline{A}_{\mp}|^2},\\
C\pm \Delta C & \equiv & \frac{|A_{\pm}|^2-|\overline{A}_{\mp}|^2}{|A_{\pm}|^2+|\overline{A}_{\mp}|^2}.
\end{eqnarray}
The measured $CP$-violation parameters for $B^0\to a_1(1260)^{\pm}\pi^{\mp}$ 
decays are summarized in Table~\ref{inputa1cp}~\cite{BTOA1PI}.

\begin{table}[h]
\begin{center}
\caption{Values of the $CP$-violation parameters used as input to the 
calculation of the bounds on $|\Delta \alpha|$~\cite{BTOA1PI}.}
\begin{tabular}{|l|c|}
\hline
$\boldsymbol{A_{CP}}$   & $-0.07\pm 0.07 \pm 0.02$ \\
\hline
$\boldsymbol{S}$        & $\phantom{-}0.37\pm 0.21 \pm 0.07$ \\
\hline
$\boldsymbol{\Delta S}$ & $-0.14\pm 0.21 \pm 0.06 $ \\
\hline
$\boldsymbol{C}$        & $-0.10\pm 0.15 \pm 0.09$ \\
\hline
$\boldsymbol{\Delta C}$ & $\phantom{-}0.26\pm 0.15\pm 0.07$ \\
\hline
\end{tabular}
\label{inputa1cp}
\end{center}
\end{table}

In analogy to the $\pi^+\pi^-$ case, where 
\begin{equation}
2\alpha_{\rm eff} = \arg\left[e^{-2i\beta}A(\bar{B}^0\to \pi^+\pi^-)A^*(B^0\to \pi^+\pi^-)\right], \nonumber
\end{equation} 
it is possible to define two quantities $\alpha^+_{\rm eff}$ 
and $\alpha^-_{\rm eff}$:
\begin{equation}
\alpha^{\pm}_{\rm eff}  \equiv \arg\left[e^{-2i\beta}\overline{A}_{\pm}A^{*}_{\pm}\right],
\end{equation}
which are related by the phase $\hat{\delta}\equiv\arg[A_+A^*_-]$
to the measurable quantities:
\begin{eqnarray}
2 \alpha^{\pm}_{\rm eff} \pm \hat{\delta} & = & \arg\left[e^{-2i\beta} 
\overline{A}_{\pm}A^{*}_{\mp}\right]\\
& = & \arcsin\frac{S\mp\Delta S}{\sqrt{1-(C\mp\Delta C)^2}}.
\end{eqnarray}
In the limit of zero penguin amplitudes, $\hat{\delta}$ coincides with the 
strong phase difference between the tree amplitudes contributing to 
$B^0\to a_1(1260)^+\pi^-$ and $B^0\to a_1(1260)^-\pi^+$ decays. 
An effective value $\alpha_{\rm eff}$ for the weak phase $\alpha$ is then
obtained as the average $\alpha_{\rm eff}=\frac{1}{2}\left(\alpha^+_{\rm eff}+\alpha^-_{\rm eff}\right)$ with an eight-fold ambiguity~\cite{GRONAUZUPAN}. 

It is possible to apply arguments based on the approximate SU(3) flavor 
symmetry to set bounds on $|\Delta \alpha|$.
The following ratios of $CP$-averaged rates of $\Delta S=0$ 
and $\Delta S=1$ transitions are calculated, that involve the same SU(3) 
flavor multiplet as $a_1(1260)$~\cite{GRONAUZUPAN}, 
such as $B^0\to a_1(1260)^- K^+$, $B^0\to K_{1A}^+ \pi^-$, $B^+\to a_1(1260)^+ K^0$, and $B^+\to K_{1A}^0 \pi^+$:
\begin{eqnarray}
R^0_+ \equiv & \frac{\overline{\lambda}^2f_{a_1}^2{\overline{\cal B}}(B^0\to K_{1A}^+ \pi^-)}{f_{K_{1A}}^2{\overline{\cal B}}(B^0\to a_1^+ \pi^-)},\\
R^+_+ \equiv & \frac{\overline{\lambda}^2f_{a_1}^2{\overline{\cal B}}(B^+\to K_{1A}^0 \pi^+)}{f_{K_{1A}}^2{\overline{\cal B}}(B^0\to a_1^+ \pi^-)},\\
R^0_- \equiv & \frac{\overline{\lambda}^2f_{\pi}^2{\overline{\cal B}}(B^0\to a_1^- K^+)}{f_{K}^2{\overline{\cal B}}(B^0\to a_1^- \pi^+)},\\
R^+_- \equiv & \frac{\overline{\lambda}^2f_{\pi}^2{\overline{\cal B}}(B^+\to a_1^+ K^0)}{f_{K}^2{\overline{\cal B}}(B^0\to a_1^- \pi^+)}.
\end{eqnarray}
The bounds are effective because the penguin contribution is CKM 
enhanced by $1/\overline{\lambda}=|V_{cs}|/|V_{cd}|$ in $\Delta S=1$ decays 
with respect to $\Delta S=0$ modes.
The following inequalities involving $(\alpha_{\rm eff}^{\pm}-\alpha)$ hold:
\begin{eqnarray} 
\cos{2(\alpha_{\rm eff}^{\pm}-\alpha)} \geq \frac{1 - 2R^0_{\pm}}{\sqrt{1-A_{CP}^{\pm 2}}}\\
\cos{2(\alpha_{\rm eff}^{\pm}-\alpha)} \geq \frac{1 - 2R^+_{\pm}}{\sqrt{1-A_{CP}^{\pm 2}}},
\end{eqnarray} 
where $A_{CP}^{\pm}$ are the direct $CP$ asymmetries 
\begin{equation}
A_{CP}^{\pm} \equiv \frac{|\overline{A}_{\pm}|^2-|A_{\pm}|^2}{|\overline{A}_{\pm}|^2+|A_{\pm}|^2}.
\end{equation}
The above relations set a constraint on $(\alpha_{\rm eff}^{\pm}-\alpha)$. 
Bounds on $|\Delta \alpha|$ are then derived from $|\Delta \alpha|\leq (|\alpha_{\rm eff}^{+}-\alpha|+|\alpha_{\rm eff}^{-}-\alpha|)/2$.

The BFs of $B\to a_1(1260) \pi$ and $B\to a_1(1260) K$ decays have been
measured in the last few years~\cite{BTOA1PI}. The measurement of the 
missing piece of input, the BFs of 
$B \to K_1(1270)\pi$ and $B\to K_1(1400)\pi$ decays, is described in 
the following section.

\subsection{$\boldsymbol{B \to K_1(1270)\pi}$, $\boldsymbol{B \to K_1(1400)\pi}$}

The $K_{1A}$ meson (the SU(3) partner of the $a_1(1260)$ meson) is a 
nearly equal superposition of the physical states $K_1(1270)$ and $K_1(1400)$.
The rates of $B \rightarrow K_{1A} \pi$ decays, which are 
experimental inputs to the calculation of the bounds on $|\Delta \alpha|$, 
must be derived 
from the measurement of the rates of $B \rightarrow K_1(1270)\pi$ and 
$B \rightarrow K_1(1400)\pi$ decays. The BFs for these
processes have recently been measured by BaBar~\cite{BTOK1PI}.
The $K_1(1270)$ and $K_1(1400)$ axial vector mesons are broad resonances 
with nearly equal masses. In the following, we will refer to them 
collectively as $K_1$. The $K_1(1270)$ and $K_1(1400)$ mesons decay 
to the same final state $K\pi\pi$, 
although through different intermediate states. However, 
since the intermediate decays 
proceed almost at threshold, the available phase spaces 
overlap and interference effects can be sizeable.
The analysis strategy relies on the reconstructed $K\pi\pi$ invariant mass 
spectrum in the $\left[1.1,1.8\right]\,{\rm GeV}$ range
to distinguish between $K_1(1270)$ and $K_1(1400)$, including 
interference effects in the signal model.

A two-resonance, six-channel $K$-matrix model is used 
to describe the resonant $K\pi\pi$ system for the signal~\cite{WA3}.
The production amplitude for channel 
$i = \{(K^*\pi)_{S-wave}, (K^*\pi)_{D-wave},$ $\rho K, K_0^*\pi, f_0K, \omega K\}$ 
is given by
\begin{equation}
F_i=e^{i \delta_i}\sum_j(\mathbf{1}-i\mathbf{K}{\boldsymbol \rho})_{ij}^{-1}\mathbf{P}_j,
\end{equation}
where $\delta_i$ are offset phases with respect to the $(K^*\pi)_S$ channel, 
\begin{equation}
K_{ij}=\frac{f_{ai}f_{aj}}{M_a-M}+\frac{f_{bi}f_{bj}}{M_b-M},
\end{equation}
and $\mathbf{P}$ is the production vector
\begin{equation}
P_i=\frac{f_{pa}f_{ai}}{M_a-M}+\frac{f_{pb}f_{bi}}{M_b-M}.
\end{equation}
The labels $a$ and $b$ refer to $K_1(1400)$ and $K_1(1270)$, respectively, 
and the indexes $i$ and $j$ refer to the final states of $K_1$ decays.
The decay  constants $f_{ai}$, $f_{bi}$,  and the $K$-matrix poles
$M_{a}$  and $M_{b}$ are real. 
The elements of the diagonal phase space matrix \mbox{\boldmath$\rho$}$(M)$
for the process $K_1 \rightarrow 3 + 4$, $3 \rightarrow 5 + 6$ have been
approximated with the form 
\begin{equation}
\rho=\frac{2\delta_{ij}}{M}\sqrt{\frac{2m^*m_4}{m^*+m_4}(M-m^*-m_4+i\Delta)},
\end{equation}
where $M$ is the mass of $K_1$,  $m_4$ is the mass of 4, 
$m^*$ is the mean mass of 3 and $\Delta$ is the half width of
$3$. 

The parameters of $\mathbf{K}$ and the offset phases $\delta_i$ are extracted
from a fit to the data collected by the WA3 experiment \cite{WA3} for the 
intensity of the 
$K\pi\pi$ channels and the relative phases. For the fit to
WA3 data a background term is included in the production vector. 

The decay constants for the $\omega K$ channel are fixed according to the 
quark model~\cite{WA3}. 
The production constants $f_{pa}$ and $f_{pb}$ are expressed in terms
of the production parameters ${{\boldsymbol \zeta}=(\vartheta,\phi)}$:
$f_{pa}\equiv \cos\vartheta$, $f_{pb}\equiv \sin\vartheta e^{i\phi}$, where
$\vartheta \in [0,\pi/2]$, $\phi \in [0,2\pi]$. 

Signal Monte Carlo (MC) samples are generated by weighting the
$(K\pi\pi)\pi$ population according to the amplitude
$\sum_{i\neq \omega K}\langle K\pi\pi| i \rangle F_i$, where the 
term $\langle K\pi\pi| i \rangle$
consists of a factor describing the angular distribution 
of the $K\pi\pi$ system resulting from $K_1$ decay, an amplitude 
for the resonant $\pi\pi$ and $K\pi$ systems, and isospin factors. 
The BF of $K_1 \to \omega K$ is
accounted for as a correction to the total selection efficiency.

The BF and the production parameters $\vartheta,\phi$ 
for neutral and charged $B$ meson decays to $K_1(1270)\pi + K_1(1400)\pi$ 
are extracted via a ML fit to the kinematic observables $m_{ES}$, $\Delta E$, 
a $\cal F$isher discriminant based on event-shape quantities, 
the $K\pi\pi$ invariant mass $m_{K\pi\pi}$ and an angular variable.  
Background from $B$ decays to $K^*(1410)\pi$ and non-resonant $B$ decays 
to $K^*\pi\pi$ and $\rho K\pi$ are taken into account as separate 
components in the fit.
The dependence of the signal probability distribution in $m_{K\pi\pi}$ 
and selection efficiencies 
on the production parameters {\boldmath$\zeta$} is described by means 
of non-parametric templates $P(m_{K\pi\pi}|\vartheta,\phi)$.

Each event is classified according to the invariant masses 
of the $\pi^+ \pi^-$  and $K^+ \pi^-$ ($K_S^0 \pi^+$)
systems in the $K_1^+$ ($K_1^0$) decay for $B^0$ ($B^+$) candidates:
events which satisfy the requirement 
$0.846 < m_{K\pi} < 0.946\,{\rm GeV}$ belong to class 1 (``$K^*$ band''); 
events not included in class 1 for which $0.500 <m_{\pi\pi} < 0.800\,{\rm GeV}$
belong to class 2 (``$\rho$ band''); all other events are rejected.

For the $B^0$ modes a likelihood scan is performed with respect to 
$\vartheta$ and $\phi$. At each point, a simultaneous fit to 
the event classes $r={1,2}$ is performed. Although for events 
in the ``$\rho$ band'' the signal to background ratio is worse
than that for events in the ``$K^*$ band'',
MC studies have shown that including those events 
in the fit helps in resolving the 
ambiguities in the determination of the parameter $\phi$.
For the $B^+$ modes, simulations show that, due to a 
less favourable signal to background ratio and increased 
background from $B$ decays, 
the analysis is not sensitive to $\phi$. A value 
$\phi=\pi$ is therefore assumed and the scan is performed only with 
respect to $\vartheta$. At each point of the scan, a fit to ``$K^*$ band''
events only is performed.

\begin{figure*}[t]
\includegraphics[width=150mm]{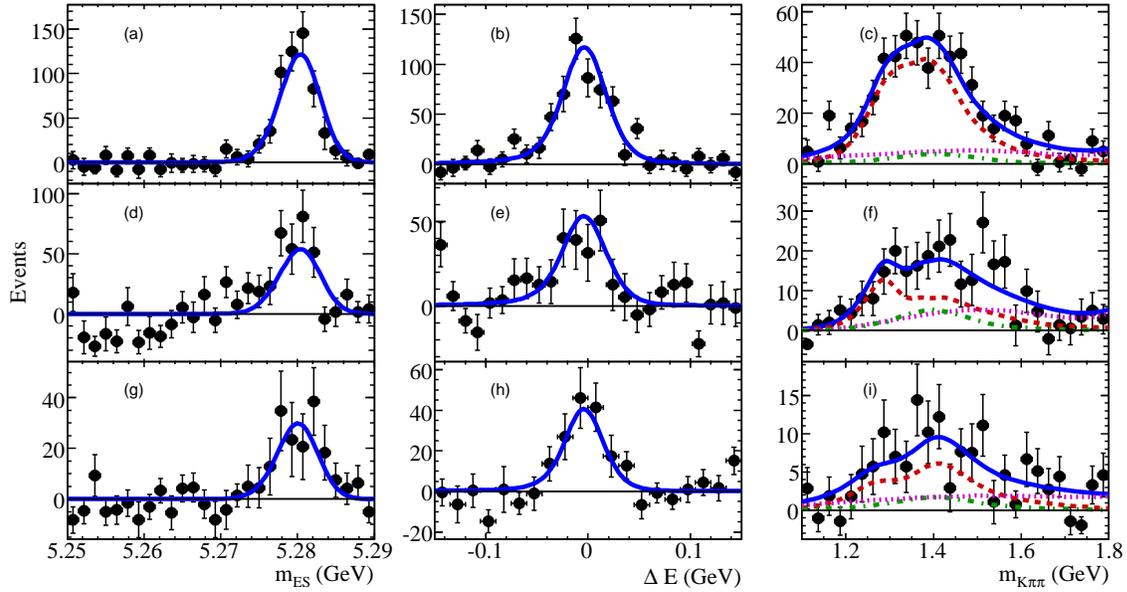}
\caption{\label{splot} 
   sPlot projections of signal onto $m_{ES}$ (left), $\Delta E$\ (center), 
   and $m_{K \pi \pi}$ (right) for $B^0$ class 1 (top), 
   $B^0$ class 2 (middle), and $B^+$ class 1 (bottom) events: 
   the points show the sums of the signal weights obtained from on-resonance 
   data.
   For $m_{ES}$ and $\Delta E$ the solid line is the signal fit function.
   For $m_{K \pi \pi}$ the solid line is
   the sum of the fit functions of the
   decay modes $K_1(1270) \pi + K_1(1400) \pi$ (dashed), 
   $K^*(1410) \pi$ (dash-dotted), and $K^*(892) \pi \pi$ (dotted), and
   the points are obtained without using information about
   resonances in the fit, \emph{i.e.}, we use only the $m_{ES}$, 
   $\Delta E$, and ${\cal F}$ variables.
}
\end{figure*}

Figure\ \ref{splot} shows the distribution of $\Delta E$, $m_{ES}$ 
and $m_{K\pi\pi}$ for the signal events obtained by the background-subtraction 
technique sPlot~\cite{SPLOT}.

\begin{figure*}[t]
\includegraphics[width=80mm]{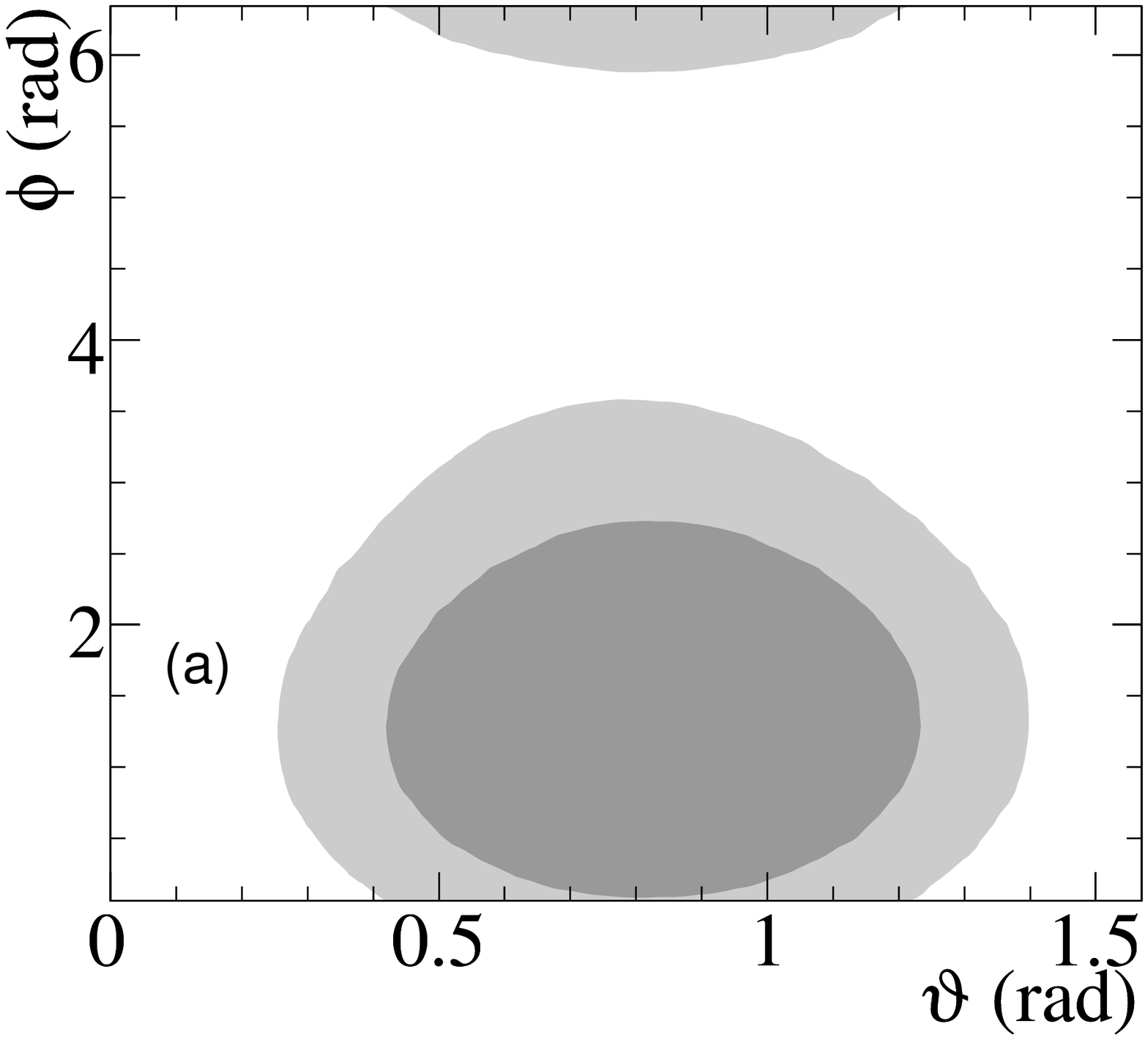}
\includegraphics[width=80mm]{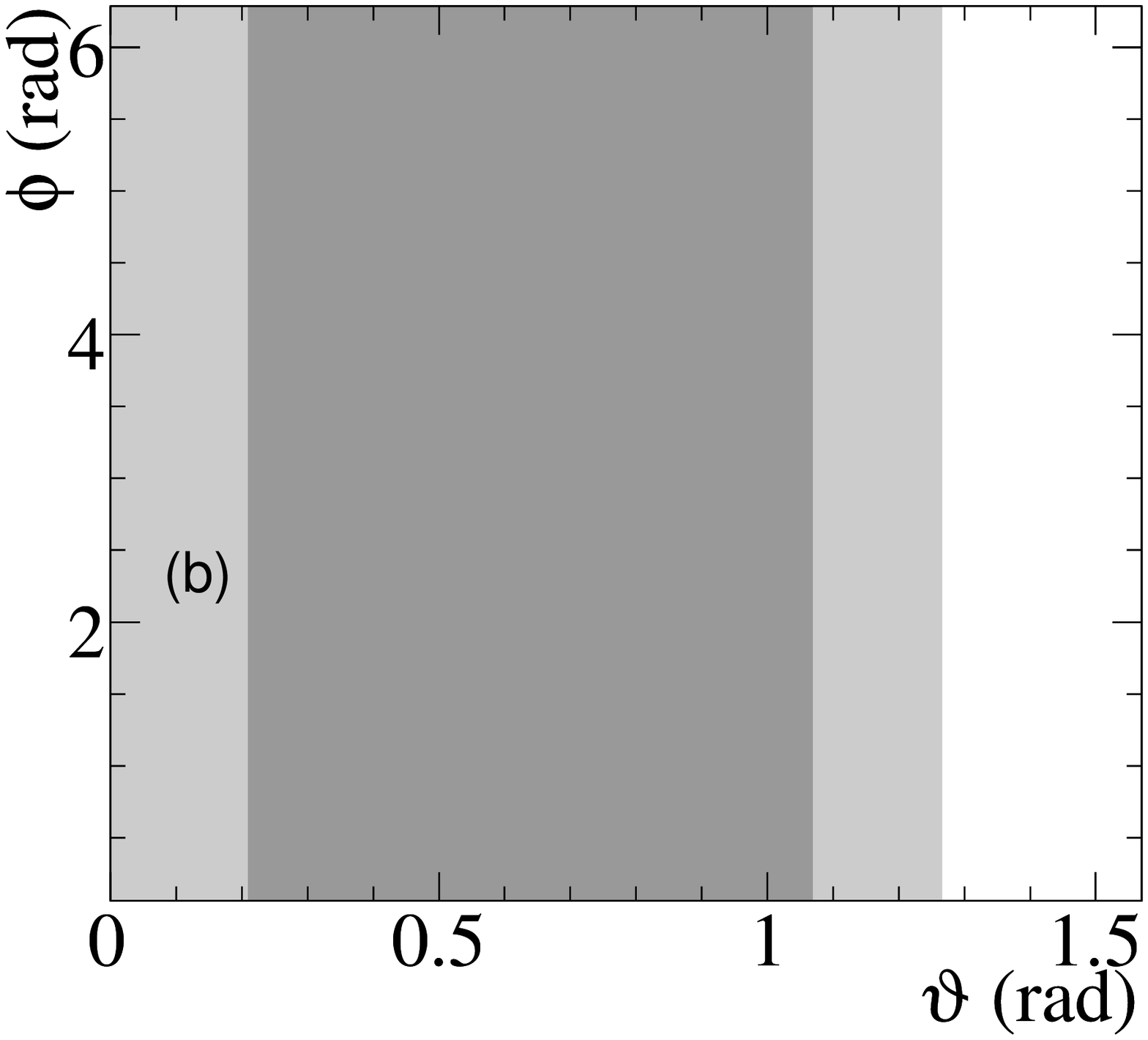}
\caption{\label{fig:nllscan} 
68\% (dark shaded zone) and 90\% (light shaded zone) probability regions 
for $\vartheta$ and $\phi$ for the (a) $B^0$ and (b) $B^+$ modes.}
\end{figure*}

The experimental two-dimensional likelihood $\mathcal{L}$ for 
$\vartheta$ and $\phi$ is convolved with a two-dimensional Gaussian 
that accounts for the systematic uncertainties.
The resulting distributions in 
$\vartheta$ and $\phi$ are shown in Fig.~\ref{fig:nllscan} 
(the 68\% and 90\% probability 
regions are shown in dark and light shading respectively, and
are defined as the regions which satisfy ${\cal {L}}(r)>{\cal {L}}_{min}$ and 
$\int_{{\cal {L}}(r)>{\cal {L}}_{min}}{\cal {L}}(\vartheta,\phi) {\rm d}\vartheta{\rm d}\phi = 68\%~~(90\%)$).

A combined signal for $B^0$ decays to $K_1(1270)^+\pi^-$ and $K_1(1400)^+\pi^-$
is observed with a significance of $7.5\sigma$, while there's evidence 
for $B^+$ decays to $K_1(1270)^0\pi^+$ and $K_1(1400)^0\pi^+$ at $3.2\sigma$.
The measured BFs are 
${\cal B}(B^0\to K_1^{+}\pi^-+K_1'^{+}\pi^-) = 31^{+8}_{-7} \times 10^{-6}$ and 
${\cal B}(B^+\to K_1^{0}\pi^++K_1'^{0}\pi^+) = 29^{+29}_{-17} \times 10^{-6}$ 
($<82\times 10^{-6}$ at 90\% probability), including systematic 
uncertainties~\cite{BTOK1PI}.

The probability distributions for the 
$B\to K_1(1270)\pi$, 
$B\to K_1(1400)\pi$, and 
$B\to K_{1A}\pi$  
BFs are derived by setting
the production parameters $(f_{pa},f_{pb})$ equal to 
$(0,e^{i\phi}\sin\vartheta)$, 
$(\cos\vartheta,0)$, and 
$(|f_{pA}|\cos\theta,-|f_{pA}|\sin\theta)$, 
respectively,
where $f_{pA}=\cos\vartheta\cos\theta - e^{i\phi}\sin\vartheta \sin\theta$ and
$\theta$ is the $K_1$ mixing angle. 
A value $\theta=72^{\circ}$ is used~\cite{BTOK1PI}.

Including systematic uncertainties the following values are 
obtained (in units of $10^{-6}$):
${\cal B}(B^0 \to K_1(1270)^{+}\pi^-) = 17^{+8}_{-11}$,~
${\cal B}(B^0 \to K_1(1400)^{+}\pi^-) = 17^{+7}_{-9}$,~
${\cal B}(B^0 \to K_{1A}^{+}\pi^-) = 14^{+9}_{-10}$,~
${\cal B}(B^+ \to K_1(1270)^{0}\pi^+) < 40$,~
${\cal B}(B^+ \to K_1(1400)^{0}\pi^+) < 39$,~
${\cal B}(B^+ \to K_{1A}^{0}\pi^+) < 36$,~
where the upper limits are evaluated at 90\% probability~\cite{BTOK1PI}.

\subsection{Extraction of $\boldsymbol{\alpha}$}
A MC technique is used to estimate a probability region 
for the bound on $|\Delta \alpha|$. The $CP$-averaged rates and $CP$-violation 
parameters participating in the estimation of the bounds are generated 
according to the experimental distributions; a summary of the experimental 
values used as input to this calculation is provided in Table~\ref{inputa1su3}.

\begin{table}[h]
\begin{center}
\caption{Summary of the $B\to a_1(1260)\pi$ and $B\to a_1(1260)K$ branching fractions (in units of $10^{-6}$) and of the form factors (in MeV) used in the calculation of $\alpha$.}
\begin{tabular}{|c|c|c|c|c|c|c|c|c|c|c|c|}
\hline
\multicolumn{4}{|c|}{$\boldsymbol{{\cal B}(a_1^{\pm}\pi^{\mp})}$ \cite{BTOA1PI}} &
\multicolumn{4}{|c|}{$\boldsymbol{{\cal B}(a_1^{-}K^+)}$ \cite{BTOA1PI}}  & 
\multicolumn{4}{|c|}{$\boldsymbol{{\cal B}(a_1^{+}K^0)}$ \cite{BTOA1PI} }\\
\hline 
\multicolumn{4}{|c|}{$33.2\pm 3.8 \pm 3.0$} & 
\multicolumn{4}{|c|}{$16.3\pm 2.9 \pm 2.3$} & 
\multicolumn{4}{|c|}{$33.2\pm 5.0 \pm 4.4$}\\
\hline
\multicolumn{3}{|c|}{$\boldsymbol{f_{\pi}}$ \cite{PDG}} &  
\multicolumn{3}{|c|}{$\boldsymbol{f_{K}}$ \cite{PDG}} & 
\multicolumn{3}{|c|}{$\boldsymbol{f_{a_1}}$ \cite{CHENG}} & 
\multicolumn{3}{|c|}{$\boldsymbol{f_{K_{1A}}}$ \cite{BLOCH}}
\\
\hline 
\multicolumn{3}{|c|}{$130.4 \pm 0.2$} &
\multicolumn{3}{|c|}{$155.5 \pm 0.9$} & 
\multicolumn{3}{|c|}{$203 \pm 18$} &
\multicolumn{3}{|c|}{$207 \pm 20$} \\ 
\hline
\end{tabular}
\label{inputa1su3}
\end{center}
\end{table}

For each set of generated values, the bound on $|\Delta \alpha|$ is evaluated.
The limits on $|\Delta \alpha|$ are obtained by counting the fraction 
of bounds within a given value and the results are 
$|\Delta \alpha|<11.1^{\circ}(13.1^{\circ})$ at 68\% (90\%) 
probability~\cite{BTOK1PI}.

The angle $\alpha$ is extracted with an eight-fold ambiguity in the range 
$[0,180]^{\circ}$. The eight solutions are $\alpha = (11 \pm 7 \pm 11)^{\circ}$, 
$\alpha = (41 \pm 7 \pm 11)^{\circ}$, $\alpha = (49 \pm 7 \pm 11)^{\circ}$, $\alpha = (79 \pm 7 \pm 11)^{\circ}$, $\alpha = (101\pm 7 \pm 11)^{\circ}$, $\alpha = (131\pm 7 \pm 11)^{\circ}$, $\alpha = (139\pm 7 \pm 11)^{\circ}$, $\alpha = (169\pm 7 \pm 11)^{\circ}$.
Assuming that the strong phase $\hat{\delta}$ is 
negligible~\cite{GRONAUZUPAN}, 
only two solutions are still allowed.
Considering only the solution consistent with the results of global CKM
fits, $\alpha = (79\pm 7\pm 11)^{\circ}$.

\section{Conclusion}

Recent updates of measurements related to the determination 
of $\alpha$ have been presented. 

The first measurement of the branching fraction of $B\to K_1\pi$ decays, 
combined with the input from the analysis of the time-dependent 
$CP$-violation asymmetries in $B^0\to a_1(1260)^{\mp}\pi^{\pm}$ decays and 
of the branching fractions of $B\to a_1(1260) K$ decays, allows to measure 
$\alpha$ in the $a_1(1260)\pi$ system. This novel determination 
of $\alpha$ is independent from, and consistent with, 
the current averages, which are based
on the analysis of the $\pi\pi$, $\rho\pi$, and $\rho\rho$ systems only.

With the new update of the $B^+\to \rho^0\rho^+$ 
branching fraction and longitudinal polarization fraction measurements, 
the determination of $\alpha$ in the $\rho\rho$ system 
has reached the unprecedented precision of $7\%$, comparable with the 
$5.3\%$ precision achieved in $\sin 2 \beta$ measurements.

In the $\pi\pi$ system, 
the updated measurement of $CP$-violating asymmetries in $B^0\to\pi^+\pi^-$
decays provides a $6.7\sigma$ evidence of $CP$ violation. 
$B^0\to\pi^0\pi^0$ decays branching fraction and direct $CP$ asymmetry 
are input to the 
isospin analysis of $B\to\pi\pi$ decays that is used to constrain 
the effect of penguin pollution on the extraction of $\alpha$. 

All the measurements described in this work have been 
performed on the final BaBar sample. Most of them are still limited
by statistics, and improvement may come from next generation, 
very high luminosity facilities. 

\begin{acknowledgments}
  I would like to thank the organizers of DPF 2009 for an interesting 
  conference and my BaBar and PEP-II collaborators for their contributions.
  I'm grateful to Vincenzo Lombardo and 
  Fernando Palombo for their support and for reviewing the manuscript.
\end{acknowledgments}

\end{document}